\input harvmac

\def\Title#1#2{\rightline{#1}\ifx\answ\bigans\nopagenumbers\pageno0\vskip1in
\else\pageno1\vskip.8in\fi \centerline{\titlefont #2}\vskip .5in}

\font\ticp=cmcsc10
 
\def\a{\rightarrow}
\def\aa{\alpha}
\def\b{\beta}
\def\ka{\kappa_{a}}
\def\kc{\kappa_{c}}
\def\l{\lambda}
\def\pp{\partial}
\def\({\left (}
\def\){\right )}
\def\[{\left [}
\def\]{\right ]}
\def\xo{\xi_1}
\def\xtw{\xi_2}
\def\xth{\xi_3}
\def\xf{\xi_4}
\def\xot{$\xi_1$}
\def\xtwt{$\xi_2$}
\def\xtht{$\xi_3$}
\def\xft{$\xi_4$}
\def\h{\hat}
\gdef\journal#1, #2, #3, 19#4#5{{\sl #1~}{\bf #2}, #3 (19#4#5)}

\lref\hsn{S. Hayward, T. Shiromizu, and K. Nakao, \journal Phys. Rev. D,
49, 5080, 1994.}
\lref\ernstm{F. Ernst, \journal J. Math. Phys., 17, 54, 1976.}
\lref\melvin{M. Melvin, \journal Phys. Lett., 8, 65, 1964.}
\lref\unstable{ S. Chandrasekar and J. Hartle, \journal Proc. R. Soc. Lond,
A384, 301, 1982; E. Poisson and W. Israel, \journal Phys. Rev., D41, 1796,
1990; A. Ori, \journal Phys. Rev. Lett., 67, 789, 1991.}
\lref\memo{F. Mellor and I. Moss, \journal Phys. Rev., D41, 403, 1990.}
\lref\bae{P. Brady and E. Poisson, \journal Class.
Quantum Grav., 9, 121, 1992.}
\lref\dggh{F. Dowker, J. Gauntlett, S. Giddings, and G. Horowitz, \journal Phys.
Rev., D50, 2662, 1994.}
\lref\dgkt{F. Dowker, J. Gauntlett, D. Kastor, and J. Traschen, \journal Phys. Rev., D49, 2909, 1994.}
\lref\ernst{F. Ernst,
\journal J. Math. Phys., 17, 515, 1976.}
\lref\kaw{W. Kinnersley and M. Walker, \journal Phys. Rev., D8, 1359, 1970.}
\lref\map{D. Markovic and E. Poisson, 
\journal Phys. Rev. Lett., 74, 1280, 1995.}
\lref\py{P. Yi, \journal Phys. Rev. Lett., 75, 382, 1995.}
\lref\wald{R. Wald,
\journal Ann. Phys., 82, 548, 1974.}
\lref\pen{R. Penrose,
\journal Rev. del Nuovo Cimento, 1, 252, 1969.}
\lref\hhr{S. Hawking, G. Horowitz, and S. Ross, \journal
Phys. Rev. D, 51, 4302, 1995.}
\lref\dgghh{F. Dowker, J. Gauntlett, G. Gibbons, and G. Horowitz, \journal
Phys. Rev. D, 53, 7115, 1996.}

\Title{\vbox{\baselineskip12pt\hbox{gr-qc/9607027}}}
{\vbox{\centerline {Tests of Cosmic Censorship in the Ernst Spacetime}}}
\centerline{\ticp Gary T. Horowitz and Harrison J. Sheinblatt}
\vskip.1in
\centerline{\sl Department of Physics, University of California,
Santa Barbara, CA 93106}
\centerline{\it gary@cosmic.physics.ucsb.edu \ \ \ hjs@cosmic.physics.ucsb.edu}

\bigskip
\centerline{\bf Abstract}
The Ernst spacetime is a solution of the Einstein-Maxwell equations describing
two charged black holes accelerating apart in a uniform electric (or
magnetic) field.  As the field approaches a critical value,
the black hole horizon appears
to touch the acceleration horizon. We show that weak cosmic censorship
cannot be violated by increasing the field past this critical value: The 
event horizon remains intact. On the other hand, strong cosmic censorship
does appear to be violated in this spacetime: For a certain range of
parameters, we find evidence that
the inner horizon is classically stable.

\Date{}

\baselineskip=16pt

\newsec{Introduction}
One of the most important open questions in classical general relativity
is Penrose's cosmic censorship hypothesis \pen,
which states that naked singularities
cannot be created by realistic physical processes. An  early test
of cosmic censorship \wald\  involved charged black holes,
which have two horizons; an inner Cauchy horizon as
well as the event horizon. These horizons coincide in the 
extremal limit
$Q=M$, and are absent for $Q>M$. Consider adding charged test particles
 with $q>m$ to a black hole with $Q<M$. If one could increase the black hole
charge $Q$ faster than its mass $M$, one could exceed the extremal limit,
and turn a black hole into a naked singularity.
However Wald showed \wald\ that cosmic censorship
could not be violated this way.
For a nearly extremal black hole, in order for a $q>m$ test particle to
reach the horizon, it must be sent in with sufficient kinetic energy so that
the net increase in the mass of the black hole exceeds the increase in the
charge.  One can interpret this result as providing
evidence that one cannot destroy the event horizon of a black hole
by forcing it to meet an inner Cauchy horizon.

Now consider a spacetime with a horizon {\it outside} the black hole
event horizon. One can ask if it
is possible in this case to destroy the event horizon  by causing it to
meet the horizon outside. One example is a black hole in de Sitter space,
which has a cosmological horizon outside the event horizon.
In this case, there is a maximum size for  a black hole set by the cosmological
constant \hsn.
If one could increase the effective cosmological constant, one could
easily violate cosmic censorship. All one would have to do is create a black
hole and then increase the cosmological constant so that the black hole
mass exceeds the maximum allowed value. Unfortunately, it is unlikely that
one can increase the effective cosmological constant without violating
a local energy condition.

Another example of a spacetime with a horizon outside the event horizon is
one describing a charged black hole uniformly accelerating in a
background magnetic field\foot{One could equally well consider electric fields,
but the form of this  field is somewhat more complicated
in the solutions we will consider.}. This
spacetime has an acceleration horizon outside the event horizon. As one
increases the magnetic field, one increases the acceleration which causes
the event horizon to move closer to the acceleration horizon.  Unlike the
effective cosmological constant, it is possible in principle
to physically increase the
strength of a background magnetic field. Can one increase the field so high
that the event horizon touches the acceleration horizon and destroys the
black hole? This is the question we wish to address in this paper. Since
the event horizon has a finite size, it would appear that it would touch
the acceleration horizon at a finite value of the acceleration and hence
a finite value of the magnetic field. Exceeding this value should turn
the black hole into a naked singularity, providing a rather `clean'
counterexample to cosmic censorship.

This question appears to be straightforward to answer since the
solution to the Einstein-Maxwell equations describing a pair of 
oppositely charged
black holes uniformly accelerating in a background magnetic field was
found by Ernst \ernst\ about twenty years ago. It was noticed in \dggh\ that the
event horizon and acceleration horizon appeared to touch at a finite value
of the magnetic field. In section 2, we review this calculation and
discuss further properties of the Ernst metric when the background field is
large. It turns out that a proper understanding of the situation requires
a special limit of the Ernst metric. This is performed in section 3,
where it is shown that event horizon never actually meets the acceleration
horizon.

In section 4 we consider a different test of cosmic censorship. A strong
form of this hypothesis states that generic solutions should be globally
hyperbolic, so that singularities are not visible even inside black holes.
This appears to be supported by studies showing that the
inner Cauchy horizon of the Reissner-Nordstr\"om solution is
unstable \unstable.
However, it has been shown that for charged black holes in de-Sitter
space, the situation is different: There are a range of parameters 
for which  the inner horizon of the Reissner-Nordstr\"om de Sitter solution 
is stable \memo.
We will argue that the same is true for accelerating black holes.
Thus even without a cosmological constant, it appears that one can violate
strong cosmic censorship. Section  5 contains some concluding comments.

\newsec{The Ernst Solution with Large Magnetic Fields}

The solution describing the background magnetic field was found by Melvin
\melvin\ and is given by
\eqn\dmelv{
\eqalign{
&ds^2=\Lambda^2\left[-dt^2+dz^2+d\rho^2\right]
+\Lambda^{-2}\rho^2d\phi^2\cr
&A_\phi={B\rho^2\over 2\Lambda}\cr
&\Lambda=1+{1\over 4}B^2\rho^2\ .\cr}
}
$A_\phi$ is the only nonzero component of the vector potential.
This solution describes a static, cylindrically symmetric flux tube. 
The flux tube has a radius of order $1/B$ and field strength of order $B$.
The total flux passing through a $z=$ constant
plane is 
$ \Phi = 4\pi /B$. Notice that as $B$ increases, the total flux decreases.

The solution describing two oppositely
charged black holes uniformly accelerating
in this background was found by Ernst \ernst\ and takes the form
\eqn\ernstmet{\eqalign{ds^2 = & {{\Lambda^2} \over{A^2 \(x - y\)^2}}
\[ G(y) dt^2 - G^{-1}(y) dy^2 + G^{-1}(x) dx^2 \] 
+ {{G(x)}\over{A^2 \(x - y\)^2 \Lambda^2}} d\phi^2 \cr 
G(\xi)= & \( 1 - \xi^2 - r_+ A \xi^3 \) \(1 + r_- A \xi \) \cr \Lambda\(x, y\) 
= & \[1 + {{1} \over{2}}qBx\]^2 + {{B^2 G(x)}\over{4A^2\(x-y\)^2}} \cr A_\phi
= &- {{2}\over{B\Lambda}} \(1 + {{1}\over{2}}qBx \). }}
with  $q^2 = r_+ r_-$.
The solution depends on four parameters $r_+, r_-, A, B$ which are
related to the mass, charge, and
acceleration of the black holes, and the background magnetic field.

The coordinate $y$ has the range $-\infty < y < x$. There is a curvature
singularity at $y = -\infty$, and $y \approx x$ describes an asymptotic
region that approaches the Melvin solution \dmelv.
It is convenient to introduce the following notation.
Let $\xi_2\le\xi_3<\xi_4$
be the three roots  of the cubic in $G$.
We also
define $\xi_1\equiv -{1/ (r_-A)}$ and choose $r_-$ so that $\xi_1 \le \xi_2$.
The function $G(\xi)$ then takes the form
\eqn\fandg{
 G(\xi)=-(r_+A)(r_-A)(\xi-\xi_1)
(\xi-\xi_2)(\xi-\xi_3)(\xi-\xi_4).}
These roots are all real only if $0< r_+A \le 2/(3\sqrt 3)$. In this case,
the surface $y = \xo$
is the inner Cauchy horizon, $y = \xtw$ is the black hole event horizon, and
$y = \xth$ is the
acceleration horizon. The limit $\xi_1 = \xi_2$ corresponds to an extreme
black hole and  was discussed in \dggh. We are interested here in
a different type of extremal limit:
If $r_+A = 2/(3\sqrt 3)$, $\xi_2 = \xi_3$, so 
the event horizon appears to coincide with the acceleration horizon.
If $r_+A > 2/(3\sqrt 3)$, the two roots 
$\xi_2 , \xi_3$ are both complex and the spacetime
has only one horizon at $y = \xo$. This spacetime describes two naked
singularities accelerating apart. The question we wish to address is
whether we can physically cause $r_+A$ to increase past the
critical value $2/(3\sqrt 3)$ by increasing the background magnetic field.

The roots $\xi_2$, $\xi_3$, and $\xi_4$ depend only on the single parameter 
$r_{+}A$. As $r_{+}A$ increases from $0$ to $2/(3 \sqrt{3})$,
$\xi_4$ decreases monotonically from $1$ to
$\sqrt{3}/2$, $\xi_3$ decreases monotonically from $-1$ to $-\sqrt{3}$, and 
$\xi_2$ increases monotonically from $-\infty$ to $-\sqrt{3}$. 
It is possible
to determine \xtwt\ and \xtht\ in terms of \xft\ only, with the result
\eqn\ttsol{\xtw = {{-\xf - \xf \(4 \xf^2 - 3\)^{1/2}}\over{2 \(1 - \xf^2\)}},
\qquad \xth = {{-\xf + \xf \(4 \xf^2 - 3\)^{1/2}}\over{2 \(1 - \xf^2\)}}.}

The coordinates ($x,\phi$) in \ernstmet\ are angular coordinates. To keep
the signature of the metric fixed, the coordinate $x$ is restricted to
the range $\xi_3 \le x \le \xi_4$ in which $G(x)$ is positive. 
One can always 
choose the range of $\phi$ so that there is no conical singularity
at one pole $x= \xi_3$ or $x= \xi_4$, but
for general choices of the parameters, there will be a singularity
at the other.
To ensure that the metric is free of conical singularities at both
poles, we must require
\eqn\nostrut{-{{G' \( \xth \)}\over{G' \( \xf \)}} =
\[ {{\Lambda\(\xth\)}\over{\Lambda\(\xf\)}} \]^2 
= \( {{1 + {{1}\over{2}} q B \xth}\over{1 + {{1}\over{2}} q B \xf}} \)^4}
and set
\eqn\delphi{\Delta \phi = {{4 \pi \Lambda^2 \( \xth \)}\over{G'\( \xth \)}}}
where we have defined $\Lambda(\xi_i) \equiv \Lambda(x=\xi_i)$. Physically,
\nostrut\ is the condition on the magnetic field which ensures that the
applied force is consistent with the given acceleration. 

We first show that the external magnetic field remains finite when
the two roots $\xi_2$ and $\xi_3$ coincide.
In this case, $G(\xi)$ has a double
zero, so $G'(\xi_3) = 0$. It then follows from \nostrut\ that
$1+{1\over 2} qB \xi_3 =0$. Since $\xi_3 = -\sqrt 3$ at the double zero,
we obtain $qB = 2/\sqrt 3$. So the product $qB$ remains finite
in this limit. However, $q$ and $B$ are the
physical black hole 
charge and asymptotic magnetic field only when the field is small.
In general, the black hole charge is
\eqn\bhcharge{\eqalign{ \h q =& {1\over 4\pi} \int F = {\Delta \phi\over 4\pi}
\[A_\phi(x=\xi_4) - A_\phi(x=\xi_3)\] \cr
=& {q \Delta \phi  (\xi_4-\xi_3)\over 4\pi (1+ {1\over 2} qB\xi_3)
(1+ {1\over 2} qB\xi_4)}}}
and the Ernst metric asymptotically approaches the Melvin solution with
parameter \dggh \dgkt
\eqn\qhbhdef{\h{B} = 
 {{B G' \( \xth \)}\over{2 \Lambda^{3/2} \( \xth \)}}}
Using \delphi, the product of the physical black hole charge and asymptotic
magnetic field is thus
\eqn\prod{ \h q \h B = {qB (\xi_4-\xi_3)\over 2 + qB\xi_4}}
Since $\xi_4 = \sqrt 3/2$ when there is a double root, $\h q \h B=1$. 
So the physical magnetic field is finite at the point when the two roots
coincide! This seems to suggest that one could violate cosmic censorship
by simply increasing the magnetic field past this finite value.

To understand this better, notice that
\nostrut\ does not fix the magnetic field uniquely, but allows
two solutions. This can be seen as follows.
Define the left hand side of \nostrut\ to be $\alpha^4$. Then, when
expressed in terms of the zeros of $G\(\xi\)$, $\alpha^4$ becomes
\eqn\alphadef{ \alpha^4 = {{\(\xth-\xtw\)\(\xth-\xo\)}
\over{\(\xf-\xtw\)\(\xf-\xo\)}}}                                                
Note that $\alpha^4 < 1$ since $\xo,\xtw < \xth < \xf$. 
Since $G(\xi)$ is independent of $B$, so is $\alpha$. We can now
solve \nostrut\ for $qB$, and there are clearly two real solutions 
\eqn\qbsoln{ qB = {{2 \( 1 \mp \alpha\)}\over{\pm \alpha \xf - \xth}}}
with $\alpha \ge 0$. Substituting this into \prod\ yields the simple
result
\eqn\qbhsoln{ \h{q} \h{B} = 1 \mp \alpha}
When $r_+A \a 2/(3\sqrt3)$ so $\xtw =\xth$, $\alpha =0$ 
and the two solutions agree. However
in the opposite limit when $r_+A \a 0$,
$\xi_1,\xi_2 \a -\infty,\
\xi_3 \a -1, \ \xi_4 \a 1$, and $\alpha \a 1$. Thus
the upper sign corresponds to the expected result
that $qB \a 0, \ \h q \h B \a 0$.  However, the lower sign corresponds to 
another branch of solutions in which $qB \a \infty$ and $\h q \h B \a 2$.

The key to understanding these new solutions is to consider the flux
of magnetic field crossing the acceleration horizon $y = \xi_3$, which 
is given by
\eqn\accflux{ \Phi = \Delta \phi \lim_{x\a\xi_3} \tilde A_\phi(x,y=\xi_3)=
{2 \Delta \phi\over B(1+ {1\over 2} qB\xi_4)}}
where $\tilde A_\phi = A_\phi - A_\phi(x=\xi_4)$ is the gauge equivalent
vector potential
which is regular on the axis $x=\xi_4$.
For small $B$, the flux is $\Phi \approx 4\pi/B$ which agrees with the flux at
infinity in the asymptotically Melvin region. This is expected since
the black holes are very small compared to the size of the flux tubes.
However, when $ q  B = O(1)$, the  black holes are comparable to the size
of the flux tube. They are also close enough so that their charges make a
significant contribution to the flux. The limit of \accflux\ when the
two roots coincide can be calculated as follows.
The range of $\phi$ \delphi\ remains finite and nonzero in this limit
since \nostrut\ implies
$\Delta \phi = 4\pi |\Lambda^2(\xi_4)/G'(\xi_4)|$. 
Since $1+ {1\over 2} qB \xi_3 \a 0$
in the limit that the two roots coincide, and $\Delta\phi$ remains finite,
it follows from \bhcharge\ that $q\a 0$. Since $qB \a 2/\sqrt 3$, the parameter
$B$ diverges. Thus, the magnetic flux crossing the acceleration horizon
goes to zero: The flux between the two oppositely charged black holes
completely cancels the external magnetic field.

Now consider the solutions with $\h q \h B >1$. This corresponds to 
the lower sign in \qbsoln\ and hence
$1+ {1\over 2} qB \xi_3 <0$. In order for
the physical charge on the black hole to have the same sign as before,
we must take the parameter $q$ negative. Since $qB$ is positive, the
parameter $B$ must also be negative. Thus the sign of the flux crossing
the acceleration horizon is reversed. This means that for the same black hole
parameters (mass, charge, and acceleration) the solution to the
constraint \nostrut\ with  $\h q \h B >1$ has a {\it smaller}
external magnetic flux than the one with $\h q \h B < 1$! This is not
a contradiction since $\h B$ is a measure of the strength of the magnetic
field on the axis, and when this is large, spacetime is highly curved, yielding
a smaller net flux. In fact, the limit $\h q \h B \a 2$ corresponds to the
two black holes becoming infinitely far apart with zero acceleration. The flux
through the acceleration horizon becomes equal in magnitude and opposite in
sign to the one at infinity. For a magnetic field $\h B = 2/\h q$ the
asymptotic flux allowed by the Melvin solution is $\Phi = 2\pi \h q$. 
Combining  this with the flux at the acceleration horizon (which is 
infinitely far away in the other direction)  one has a net outward
flux of $\Phi = 4\pi
\h q$ which is just the flux of a charge $\h q$. In other words,  the limit
$\h q \h B \a 2$ corresponds to turning off the external magnetic field
completely! One has compressed the magnetic flux from a charged black hole
into two tubes so that the gravitational attraction exactly balances the
magnetic pressure resulting in a static configuration.

This shows that the branch of solutions with $1<\h q\h B<2$ does not
correspond to physically increasing the external magnetic flux. To understand
whether one can violate cosmic censorship this way,
we must study the limit $\h q \h B \a 1$
more closely.
We will do this in the next section. For completeness, we will also
study the limit $\h q \h B \a 2$.

\newsec{Limiting Solutions}

We wish to investigate the limit of the  Ernst solution
\ernstmet\ as $\h q \h B \a 1$ and 
 $ \h q \h B \a 2$
keeping the  black hole unchanged. It is not completely obvious what it
means to say that black holes in two different spacetimes are the same.
One cannot fix the mass,  since there is
no completely satisfactory definition of quasilocal mass for a nonspherical
spacetime. Instead, we will keep the black hole charge
and horizon area fixed. From \ernstmet\
the horizon area is
\eqn\area{ Area = {\Delta \phi (\xi_4- \xi_3)\over
A^2 (\xi_3- \xi_2)(\xi_4- \xi_2)}}
with $\Delta \phi $ given by \delphi. The limits $\h q \h B \a 1,\
\h q \h B \a 2$ must be taken carefully,
since we have seen that the first requires $q \a 0, \ B \a \infty$, while
the second requires $qB \a \infty$.

	To investigate the limit $\h q \h B \a 1$, we set
\eqn\xfdef{\xf = {{\sqrt{3}}\over{2}} + \epsilon^2.}
Using \ttsol\ one finds that to first order in $\epsilon$,
\xtwt\ and \xtht\ are given by
\eqn\xtwthlim{\xi_{2,3} =
-\sqrt{3} \mp {{\gamma}\over{2}} \epsilon \qquad \gamma \equiv 3^{3/4} 4.}
The behavior of \xot\ is not determined since it depends only on $r_-A$,
which is independent
of $r_+A$.  Thus we are free to define its behavior for small $\epsilon$
in terms of a parameter
$\beta\ge \gamma$:
\eqn\betdef{\xth - \xo = \beta \epsilon.}
With these definitions, $\alpha$ can be determined from \alphadef\
to be $\alpha = \[4 \beta\gamma /{27}\]^{1/4} \epsilon^{1/2}$.  Thus, in this 
limit
\eqn\qbeplim{qB = {{2}\over{\sqrt{3}}} - \[{4 \beta \gamma}\over{3}\]^{1/4}
\epsilon^{1/2}, \qquad \h{q} \h{B} = 1 - \[4\beta\gamma\over{27} \]^{1/4} 
\epsilon^{1/2}}
where the sign has been chosen to correspond to weak magnetic field in
the limit of small acceleration.

We have already seen that the range of $\phi$ \delphi\ remains finite
and nonzero in the limit
$\epsilon \a 0$. In order for the
black hole area \area\ to remain finite and nonzero, we clearly need to rescale
the parameter A
\eqn\rsa{ A = {\h A \over \epsilon^{1/2}}.}
Since $r_+ A$ and $r_-A$ both have nonzero limits, this implies
$q = \sqrt{r_+ r_-} = O(\epsilon^{1/2})$, so that the physical charge
\bhcharge\  also remains finite\foot{One can add higher order corrections to
ensure that the black hole area and charge are actually constant for finite
$\epsilon$. The above discussion will be sufficient to determine the limiting
solution.}.

In order to obtain a well defined metric in the limit $\epsilon \a 0$,
we clearly have to introduce new coordinates. These coordinates
can be derived by examining the asymptotic form of the solution and using
the fact that the Melvin metric \dmelv\    remains well defined for all
magnetic fields \foot{A different choice of coordinates which does not
preserve the asymptotic form of the solution in the limit
was given in \dgkt.}. The result is
\eqn\coordresc{\eqalign{x = & \xth + \epsilon \h{x} \cr 
y = & \xth + \epsilon \h{y} \cr t = & \eta \epsilon^{-2}}}
So the new coordinates $\h{x}$, $\h{y}$, and $\eta$ are just 
rescaled versions of
the old ones. 
Since $y \rightarrow x = \xth$ describes spatial infinity  
this rescaling ``opens up'' a small
region about the symmetry axis pointing toward spatial infinity.
In terms of these new coordinates, the metric functions become
\eqn\epsdep{\eqalign{G(x)
= & \epsilon^3 \h{G} \( \h{x} \),
\qquad \h{G} \( \h{x} \) = {{1}\over{\sqrt{3}}} \h{x} \( \h{x} + \gamma \) \( \h{x} + \beta \) \cr 
\Lambda \( x,y \) = & \epsilon \h{\Lambda} \( \h{x}, \h{y} \),
\qquad \h{\Lambda} \( \h{x},\h{y} \) = {{1}\over{2}} \sqrt{3 \beta \gamma} + {{3}\over{2}} {{\h{G} \( \h{x} \)}\over{\( \h{x} - \h{y} \)^2}}}}

The Ernst metric \ernstmet\ now has a well behaved limit as $\epsilon \a 0$
\eqn\limmeto{ ds^2 = {{\h{\Lambda}^2}\over{\h{A}^2 \( \h{x} - \h{y} \)^2}}
\[ \h{G} \( \h{y} \) d\eta^2 - \h{G}^{-1} \( \h{y} \) d\h{y}^2 +
\h{G}^{-1} \( \h{x} \) d\h{x}^2 \] + {{\h{G} \( \h{x} \) d\phi^2}\over{\h{A}^2 \( \h{x} - \h{y} \)^2 \h{\Lambda}^2}}}
and the limiting vector potential is
\eqn\limmag{A_\phi = -{(\beta\gamma/3)^{1/4}\over \h A \h \Lambda}}
 There are two free parameters $\h A$ and $\beta$ ($\gamma$ is a fixed
 constant \xtwthlim) which is what one expects, since the original Ernst
 solution had three free parameters and we have fixed the magnetic field.
 In this metric, $0 \le \h x < \infty$ with one pole of the
sphere at $\h x =0$ and the other at $\h x = \infty$. It is clear from 
\coordresc\
that as $\epsilon \rightarrow 0$, $\h{x}$ must approach infinity to make
$x \rightarrow \xf$ since $\xf - \xth$ is finite.  There is no conical
singularity at either pole of the sphere. 
The coordinate transformation $r = \h{x}^{-1/2}$,
$\phi = {{3 \sqrt{3}}\over{2}} \tilde{\phi}$ applied to the asymptotic form
of the $\(\h{x}, \phi \)$
metric,
\eqn\xhpmet{{{3 \sqrt{3}}\over{4 \h{A}^2}} {{d\h{x}^2}\over{\h{x}^3}} 
+ {{4 d\phi^2}\over{3 \sqrt{3} \h{A}^2 \h{x}}}}
gives
\eqn\rptilmet{{{3 \sqrt{3}}\over{\h{A}^2}} \( dr^2 + r^2 d\tilde{\phi}^2 \)}
which is regular as $r \rightarrow 0$.

The range of $\h y$ is $-\infty < \h y < \h x$ as before. The surface
$\h y = - \beta$ is the inner Cauchy horizon, $\h y = - \gamma$ is the
event horizon, and $\h y =0$ is the acceleration horizon. Thus despite the
fact that the roots $\xi_2$ and $\xi_3$ coalesce in the original coordinates,
the
black hole horizon and acceleration horizon {\it do not coincide} in the
limiting spacetime. It was noticed earlier \dgkt\ that the proper
distance between these two horizons does not go to zero when the two
roots coincide. However, this is also true for the inner and outer horizons
of the Reissner-Nordstr\"om
solution. The point is that there are different ways to take the extremal
limit. Physically, one wants to take the limit in such a way that
preserves the asymptotic behavior of the solution. In the 
Reissner-Nordstr\"om case, this implies that the two horizons
coincide in the extremal limit, while in the Ernst metric, we have
found that they do not.

We now consider the second limit $\h q \h B \a 2$. This corresponds to
the limit of small $r_+A $, but with the lower sign in
\qbhsoln. Since the analysis is similar to that above, our discussion will
be brief. Define $\delta < 0$ and $\beta \ge 1$ by
\eqn\dbdef{\delta = 1/ \xtw, \qquad \xo = \beta \xtw = \beta/\delta}
We are interested in the regime $|\delta|\ll 1$, where $\delta \approx -r_+A $.
Then to first order in $\delta$, $\xi_3 = -1 + {1\over 2} \delta$ and
$\xi_4 = 1+ {1\over 2} \delta$
so that
\eqn\oqb{qB = {{-8\beta}\over{\delta \( 1 + 3\beta \)}}.}
Requiring that the solution approach the Melvin metric asymptotically, and the
physical charge and the area of the event horizon remain
finite and nonzero as $\delta \a 0$ motivates the following definitions:
\eqn\secdefs{\eqalign{c \equiv {1\over{2}} qB\delta,
\quad A = & -{c^2\over{\delta}}\h{A},  \cr 
y = {\h y\over{\delta}}, \quad t = -\delta \h{t},
\quad x = & -\cos{\theta}, \quad \phi = c^4\h{\phi}/\delta^4}}
Since $\xi_3 \le x \le \xi_4$, in the limit $r_+ A \a 0$, we have
$G(x) = 1-x^2$. The other metric functions become
\eqn\implic{\eqalign{ \Lambda = {{c^2}\over{\delta^2}}\h{\Lambda},
\qquad & \h{\Lambda} = {\cos^2{\theta}} + {\beta\over{\h{y}^2}}{\sin^2{\theta}},
\cr G\(y\) = {1\over{\delta^2}}\h{G}\(\h{y}\), 
\qquad & \h{G}\(\h{y}\) = - {1\over{\beta}}\h{y}^2\(\h{y} - \beta \)
\(\h{y} - 1\)}}
Defining $r=1/\h{y}$ and $G(r) = \(1 - {1\over{r}}\)\(1 - {1\over{\beta r}}\)$
so that $G(\h{y})/\h{y}^2 \a - G(r)$, the limit of the Ernst solution 
as $\delta \a 0$ is
\eqn\finmet{\eqalign{ ds^2 =&
{{\h{\Lambda}^2}\over{\h{A}^2}} \[- G\(r\) d\h{t}^2 
+ {{dr^2}\over{G\(r\)}} + r^2 d\theta^2 + 
{{r^2 {\sin^2{\theta}}}\over{\h{\Lambda}^4}}d\h{\phi}^2 \] \cr
A_{\h \phi} =&-{\cos\theta\over \h A \sqrt\beta \h \Lambda}}}
Near the poles, $\h{\Lambda} \simeq 1$, so there are no conical singularities. 
There is a curvature singularity at $r=0$, an inner horizon at $r = 1/\beta$,
and an event horizon at $r=1$. The acceleration horizon ($\h y =0$)
has moved off
to infinity ($r = \infty$). The solution \finmet\ describes a single 
magnetically charged 
black hole whose magnetic flux is confined to two flux tubes.
It can alternatively be obtained as a limit of another class of solutions
describing a single black hole
in a background magnetic field \ernstm.

\newsec{Violation of Strong Cosmic Censorship in  Charged C-metric}

We now turn to another test of cosmic censorship involving the stability
of inner horizons. Since the background magnetic field will not play
an essential role here, we will set it to zero. The Ernst solution 
\ernstmet\ with $B=0$ is just the charged C-metric.
We will apply a general argument which indicates the stability
of Cauchy horizons \bae\ to this spacetime.
Consider an ingoing flux of null radiation with a finite energy density
at the acceleration horizon, as measured by a freely falling observer.
If the
energy density of the radiation at the Cauchy horizon, again measured by
a freely-falling observer, remains finite, then the Cauchy horizon is likely
to be stable. (This has been confirmed in the case of Reissner-Nordstr\"om
black holes in de Sitter space \bae.) This
condition will lead to the conclusion that the Cauchy horizon is probably
stable whenever the surface gravity of the Cauchy horizon, $\kc$, is
less than the surface gravity of the acceleration horizon, $\ka$.
We will see that for the charged C-metric solution, ranges of the
parameters exist for which this condition is satisfied.

	The charged C-metric in retarded coordinates is \kaw
\eqn\metric{\eqalign{ ds^2 & = 
H du^2 + 2 du dr + 2 A r^2 du dx - r^2 \( G^{-1} dx^2 +
G d\phi^2 \) \cr 
H & = -A^2 r^2 G\(x - {1\over Ar} \) \cr
G & = 1 - x^2 - 2 m A x^3 - e^2 A^2 x^4.}}
Note that the form for $G(x)$ is different than in the previous sections;
this is merely a coordinate gauge freedom.  There are still four roots of 
$G$:  $\xi_{1}$, $\xi_{2}$, $\xi_{3}$, and $\xi_{4}$.
Again, $x$ must be restricted to lie between $\xi_{3}$ and $\xi_{4}$ to keep the
correct signature of the metric.  $u$ is a null coordinate ranging from
$-\infty$ to $\infty$, and $r$ is a radial coordinate ranging from $0$ to
$\infty$. As before, the horizons occur at the zeros of $G$, with
the Cauchy horizon at $\xi_1$, the black hole horizon at $\xi_2$ and
the acceleration horizon at $\xi_3$. At these horizons, $H$
(the norm of the boost killing field $\pp / \pp u$) vanishes.
In addition, the null surface at the
boundary of this coordinate system, $u \rightarrow \infty$, consists of
part of the acceleration horizon and part of the Cauchy horizon.

Consider an
ingoing null flux of radiation with the stress-energy tensor
\eqn\emten{T_{\aa\b} = \[ L\( u \)/\( 4 \pi r^2 \) \] l_{\aa} l_{\b}}
 where $l_{\aa}$ is
tangent to ingoing null geodesics.  
The general ingoing  null geodesic can 
be shown to be \kaw\ 
\eqn\nultan{\eqalign{ l^{\aa} & = \( H^{-1}\(E+R \),-R-AP,{P \over r^2},{J_{z} \over r^2 G} \) \cr
R & = \( E^2 - J^2 H/r^2 \)^{1/2} \cr
P & = \( G J^2 - J_{z}^2 \)^{1/2}}}
where $E$, $J_{z}$ and $J$ are constants of the
motion.  Since we are interested in radial null geodesics, we 
set $J$ and $J_{z}$
to zero in \nultan\ to give
\eqn\radnultan{l^{\aa} = \( 2E/H, -E, 0, 0\).}

	The energy density seen by a freely-falling observer with four-velocity
$v^{\aa}$ near the acceleration horizon is $\rho_{a} = 
T_{\aa\b} v^{\aa} v^{\b} =
\[ L\( u\)/ \( 4 \pi r^2 \) \] \(l_{\aa} v^{\aa} \)^{2}$.
A family of timelike geodesics near a horizon is given by 
\eqn\tanvec{\eqalign{ v^{\aa } \( \l \) = \( \right. \dot{u} \( \l \) , \dot{r} &
\( \l \) , \dot{x} \( \l \) , \left. 0 \) \cr 
\dot{u} \( \l \) \approx {1 \over \kappa_{\xi} \l }, \quad \dot{r} \( \l
\) \approx r_{0} + r_{1} & \l, \quad \dot{x} \( \l \) \approx x_{0} +
x_{1} \l }}
where $ r_{1}, x_{0}, x_{1}$ label the different geodesics,
$\l$ is an affine parameter which goes to zero on the horizon, 
$\kappa_{\xi}$ is the surface
gravity of the horizon, and $r_0$ is determined by  $x_{0}-{1 \over
Ar_{0}} = \xi$, a zero of $G$. These conditions only fix $\lambda$ up to
a constant rescaling. The surface gravity is also ambiguous up to a
constant rescaling since it depends on the timelike Killing field which
does not have a canonical normalization at infinity. We will take
the Killing field to be simply $\psi^{\aa} = \(1, 0, 0, 0\)$ in our coordinates.
Then the surface gravity of the horizon associated with root $\xi$ of $G$,
is
\eqn\surfgrav{\kappa_{\xi} = {1 \over 2} A \left|{\left. {d G\( a \)
\over d a} \right|_{a=\xi} }\right| .}
We will only need the ratio of two different 
surface gravities which is independent of the ambiguity in their definition.

$(\lambda, r_{1}, x_{0}, x_{1})$ are good coordinates near the horizon.  In 
terms of these coordinates, $H$ and its derivatives take the form:
\eqn\fundefs{\eqalign{ H \approx -2 \kappa \( Ar_{0}^{2}x_{1} +
r_{1} \) \l, \quad {\pp H \over \pp r} \approx & -2 \kappa, \quad {\pp H \over
\pp x} \approx -2A \kappa r_{0}^2 }}
One can easily verify that $v_{\aa}v^{\aa} = -1$ on the horizons
and that the form for $v^{\aa}$ given in \tanvec\
is consistent with the geodesic equations
near a horizon.

Integrating the equation for $\dot{u}$ \tanvec\ shows that near the
horizon ${1 \over \l} = e^{\kappa u}$.  Since $l_{\aa}v^{\aa} \propto
{1 \over \l}$, it follows that the energy density at the acceleration
horizon is
\eqn\endensa{\rho_{a} \propto L\( u \) e^{2 \ka u}}
and the energy density at the Cauchy horizon is
\eqn\endensc{\rho_{c} \propto L\( u \) e^{2 \kc u}.}

	The requirement that $\rho_{a}$ be finite gives the condition on
$L\( u \)$ that
\eqn\lkcond{L\( u \) = K\( u \) e^{-2 \ka u}}
where
\eqn\kcond{\lim_{u \a \infty} K\( u \) = K_{\infty} \ne 0.}
Plugging this form for $L\( u \)$ into the equation for $\rho_{c}$ gives
\eqn\result{\rho_{c} \propto K\( u \) e^{2\( \kc - \ka \) u.}}
Clearly this expression is finite whenever $\kc < \ka$.  This condition can
be met for a range of parameters.  $G(x)$ can be written in the form 
\eqn\gform{ G(x) = -e^2 A^2 \(x - \xo\) \(x - \xtw\) \(x - \xth\) \(x - \xf\) .}
The ratio of the surface gravities is then
\eqn\sgrat{ {{\ka}\over{\kc}} = {{G'\(\xth\)}\over{G'\(\xo\)}} = {{\(\xth - \xtw\)\(\xf - \xth\)}\over{\(\xtw - \xo\)\(\xf - \xo\)}}.}
 From the form of $G(x)$ in \metric\ it is clear that by increasing $e$, thus
making the quartic term more negative, the maximum of $G$ which occurs at a 
negative value of $x$ must decrease.  This will cause the lower two zeroes
of $G$, $\xo$ and $\xtw$, to approach one another.  Thus, by increasing $e$
the ratio of surface
gravities can become arbitrarily large. This corresponds to black holes which
are near extremality. Therefore nearly extremal accelerating black holes
are likely to have stable Cauchy horizons.

\newsec{Discussion}

We have considered two oppositely charged black holes  uniformly accelerating
in a background magnetic field. This situation is described by
the Ernst solution of Einstein-Maxwell theory. To test cosmic censorship,
we studied the effect of
increasing the external magnetic field. In the standard coordinates,
it appeared that the event horizon approached the acceleration horizon
at a finite value of the background field. We showed that this is not
the case: When the limit is taken carefully, one finds that these two
horizons remain a finite distance apart. 
Thus, the situation is different
from the Reissner-Nordstr\"om de Sitter metric, in which the event horizon
can be made to coincide with the  cosmological horizon at a finite
value of the cosmological constant. We have also found another branch of
Ernst solutions  with larger values of the  asymptotic
magnetic field, but argued that this corresponds to decreasing the
external flux and concentrating the flux due to the charged black holes.

Although weak cosmic censorship cannot be violated using the Ernst solutions, 
one is left with the physical question of what happens if one continues
to increase the external magnetic flux. The result is apparently not
described by any metric of the Ernst form. This is plausible if
one recalls that the Ernst metric has zero energy relative to the
asymptotic magnetic field. This is a consequence
of the boost symmetry\foot{Not only does the usual energy associated with
an asymptotic time translation vanish, but so does
the boost energy associated with the timelike Killing field \hhr.}.
There must exist other solutions describing oppositely charged black holes
in magnetic fields with both positive and negative energy. In some cases
the black holes will collide, and in others they will expand apart. In
particular, there must exist a static solution analogous to the ones 
discussed in \dgghh. One will have to find and
study the solutions where the black holes
collide to determine if weak cosmic censorship can be violated in this case.

Within the context of the Ernst solutions, we have
argued that strong cosmic censorship is violated.  If one sends in a flux 
of null radiation which has finite energy density as measured by a 
freely-falling observer near the acceleration horizon, one finds that 
an open set of parameters exist for which the energy density as measured by 
freely-falling observers near the Cauchy horizon remains finite.  Thus,
unlike the Cauchy horizon of the Reissner-Nordstr\"om spacetime, the Cauchy 
horizon of the C-metric does not have a generic infinite blue-shift 
instability.

It may then be asked if cosmic censorship is violated semi-classically.
Arguments have been made for the Reissner-Nordstr\"om-deSitter spacetime,
which has a similar causal structure to that of Ernst, that while 
configurations with stable Cauchy horizons exist classically,
only a 
set of measure zero exist semi-classically \map.  
Near a horizon of the Ernst spacetime
the scalar wave equation decouples in terms of Rindler modes and angular 
eigenfunctions \py, but in general the lack of spherical symmetry makes 
this problem difficult.  It seems likely, however, that the Ernst 
Cauchy horizon will exhibit the same semiclassical instability as 
the Reissner-Nordstr\"om-deSitter solution.

\vskip 1cm
\centerline{\bf Acknowledgements}
This work was begun at the Newton Institute, Cambridge.  
It was supported in part by NSF Grant PHY95-07065.
\listrefs

\end